\documentclass[twocolumn]{aa} 
\usepackage{graphicx}
\begin{document}
\title{The absolute motion of the peculiar cluster NGC~6791.\thanks{Based
on observations with the NASA/ESA {\it Hubble Space Telescope},
obtained at the Space Telescope Science Institute, which is operated
by AURA, Inc., under NASA contract NAS 5-26555, under programs
GO-9815, and GO-10471.}}

       \author{Luigi R.\ Bedin\inst{1},
               Giampaolo Piotto\inst{2},
               Giovanni Carraro\inst{2,3},
               Ivan R.\ King\inst{4},
               Jay Anderson\inst{5}.
              }


       \institute{European Southern Observatory, Garching,
             Karl-Schwarzschild-Str.\ 2, D-85748, EU\\
                 \email{lbedin@eso.org}
             \and
   	      Dip.\ di Astronomia, Univ.\ degli studi di Padova,
             Padova, vic.\ Osservatorio 2, I-35122, EU\\
                  \email{piotto@pd.astro.it}
             \and
                  Andes Fellow, Departamento de Astron\'omia, Universidad
    	 de Chile, Casilla 36-d, Santiago, Chile\\
                  \email{gcarraro@das.uchile.cl}
    	 \and
             Dept.\ of Astronomy, Univ.\ of Washington,
             Box 351580, Seattle, WA 98195-1580, USA\\
                 \email{king@astro.washington.edu}
    	 \and
             Dept.\  of Physics  and Astronomy,  Mail Stop  108, Rice
             University,  6100  Main  Street,  Houston,  TX  77005,  USA\\
                 \email{jay@eeyore.rice.edu}
                 }

       \date{Received YY YYYYY 200Y / Accepted XX XXXXX 200X}

\abstract{
We present  improved values  of the three  components of  the absolute
space velocity of the open cluster NGC 6791.
One  {\sl  HST} ACS/WFC  field  with  two-epoch observations  provides
astrometric measurements of objects  in a field containing the cluster
center.  Identification  of 60  background galaxies with  sharp nuclei
allows us  to determine an  absolute reference point, and  measure the
absolute  proper   motion  of  the  cluster.    We  find  ($\mu_\alpha
\cos{\delta}$,  $\mu_\delta$)$_{\rm J2000.0}$  $=$ ($-0.57  \pm 0.13$,
$-2.45 \pm  0.12$) mas yr$^{-1}$, and  adopt $V_{\rm rad}  = -47.1 \pm
0.7$ km s$^{-1}$ from the average of the published values.
Assuming a Galactic potential, we  calculate the Galactic orbit of the
cluster  for  various  assumed  distances,  and  briefly  discuss  the
implications on the nature and the origin of this peculiar cluster.
       }
       \titlerunning{Absolute motion of NGC~6791}
       \authorrunning{Bedin et al.}
       \maketitle

%
\section{Introduction}
%

NGC 6791 is  a unique object in our Galaxy.   Usually classified as an
open cluster, it has a  number of peculiarities which leave its origin
and its  nature quite  enigmatic.  It is  more massive (at  least 4000
$m_\odot$), more metal-rich ([Fe/H] $\sim +0.4$, Carraro et al.\ 2006,
Gratton et al.\ 2006), and older  ($\approx9 $ Gyr, King et al.\ 2005)
than most known open clusters.   In contrast with other open clusters,
which lie close to the Galactic  plane, its distance of $\sim$ 4000 pc
(King et al.\  2005) and Galactic latitude of  11$^\circ$ put NGC 6791
$\sim$  1 kpc above  the plane.   The cluster  is also  anomalous with
respect  to  the radial  abundance  gradient  and the  age-metallicity
relation of  the Galactic  disk.  (See discussion  in Carraro  et al.\
2006.)

All of  these peculiarities  make NGC 6791  both an interesting  and a
challenging object, and they stimulated  us to undertake a study of it
with  deep {\sl HST}  imaging.  In  such {\sl  HST} programs  we have,
whenever possible, included a second epoch of observation, in order to
use  proper motions  to separate  cluster stars  from the  field.  The
second epoch also allows us, here,  to study the motion of the cluster
itself.

We  have  already  published  two  papers  based  on  the  first-epoch
observations alone.   The first of them reported  the discovery (Bedin
et  al.\ 2005)  of an  anomalous  white dwarf  cooling sequence.   The
second paper (King et al.\ 2005) was devoted to the main sequence, and
included  a preliminary  mass function,  which we  found to  be rather
flat.

Along with  our proper motion,  the availability in the  literature of
radial  velocities  for a  number  of  cluster  members allows  us  to
determine all three components of  the absolute motion of NGC 6791 and
to infer some  properties of its orbit, which will  shed some light on
the possible origin  of this object.  A detailed  study of the cluster
main sequence  (down to the  hydrogen-burning limit) and of  the white
dwarf cooling sequence will be presented in forthcoming papers.

%
\section{Observations, data reduction, and proper motions}
%

All of  the observations used  for the proper-motion  measurement come
from our {\sl HST} programs GO-9815 and GO-10471 (PI King), which were
separated by $\sim 2$ years.  For precise astrometric measurements and
a more accurate assessment of the errors, we had taken particular care
to   dither   our  images   properly,   with   both  whole-pixel   and
fractional-pixel  offsets  (following  the  general recipes  given  in
Anderson  \&  King   2000).   Table~\ref{obs}  describes  the  ACS/WFC
observations used in this work.
The first of the two programs  also included images in the F606W band,
but we  used only  F814W images  for our proper  motions, in  order to
avoid any possible filter-dependent systematic errors.

We measured  positions and fluxes for  every star in  every F814W {\tt
\_FLT}  exposure,  using  library  effective  PSFs  and  the  software
programs documented in  Anderson \& King (2006).  We  then generated a
master list  of all  the stars, and  collated all the  observations of
each star.   As in Bedin et  al.\ (2003), we used  the best distortion
corrections  available  (Anderson  2002,  2005)  to  correct  the  raw
positions that we had measured from the {\tt \_FLT} exposures.

We carefully constructed the reference frame, as follows.  We measured
a  simple centroid position  for each  bright star  in the  first {\tt
\_DRZ}  image, and  found  by least  squares  a linear  transformation
between those  positions and  the positions of  the same stars  in the
corresponding {\tt  \_FLT} image.   We then rotated  the frame  of the
{\tt  \_FLT} image  so that  the $y$  axis points  exactly  north, and
rescaled it  to agree with  the pixel size  of the {\tt  \_DRZ} image,
which is 50 mas/pixel.  This  is our reference frame.  From the header
of the {\tt \_DRZ} image we got  the R.A.\ and dec of the point in our
reference frame that corresponds to  the position given in the header;
this allows  us to combine  the internal accuracy of  our PSF-measured
positions  with the  absolute orientation  and scale  information from
calibrated pipeline products.

As   reference   stars   we   identified  cluster   members   in   the
color-magnitude   diagram,  and   used  only   those  stars   for  the
transformation  from each  exposure  into the  reference frame.   (See
Bedin  et al.\ 2003  for details.)   We thus  ensured that  the proper
motions are measured relative to  the bulk cluster motion.  Carraro et
al.\ (2006)  give an  estimate of  2.2 $\pm$ 0.4  km s$^{-1}$  for the
internal dispersion  of radial  velocities; for a  distance of  4 kpc,
this  corresponds to  a  proper-motion dispersion  of  $\sim$ 0.1  mas
yr$^{-1}$ ($\sim$ 0.005 ACS/WFC pixel  in two years).  This means that
the relative motions of the cluster stars should all be zero to within
the measurement errors.
We  iteratively removed  from  the  member list  some  stars that  had
field-star-type motions, even though their colors placed them near the
fiducial  cluster   sequence.   Field-star  proper   motions  will  be
discussed very briefly at the end of this section.

Finally,  in  order  to  minimize  the influence  of  any  uncorrected
distortion on  transformations into the  reference frame, we  used for
each  object a  local transformation  based on  the nearest  $\sim$ 50
well-measured  cluster stars.   With all  these precautions,  we found
that  for stars  with  $>$500 DN  in  their brightest  pixel we  could
measure positions  in a  single image with  an error $<$0.05  pixel in
each coordinate.  We  note that the relatively high  background in our
images ($\sim$ 65 DN)  makes the astrometric effect of charge-transfer
efficiency negligible, and also that the red-halo effect that disturbs
ACS/HRC  images is  negligible  in  WFC images  taken  with the  F814W
filter.

A visual  inspection of the  images reveals many  background galaxies.
Since 60 of these show  a point-like nucleus, we used our ePSF-fitting
procedure  to measure  positions for  them also.   We  transformed the
position of each  galaxy into the reference frame  using the same kind
of transformations that we used  for the stars.  With $N$ observations
for each galaxy at each epoch, we calculated the random error for each
galaxy  as  $\sigma  /\sqrt{N-1}$,   where  $\sigma$  comes  from  the
agreement  among  the  independent  measurements.  We  also  added  in
quadrature the error in the  transformation, based on the residuals of
the stars used to compute the transformation.  As expected, the errors
in  multiple measurements  of galaxy  positions are  several  times as
large as typical errors in  star positions, and depend strongly on the
galaxy morphology.  Nevertheless, on  average these galaxies provide a
good reference frame for the measurement of absolute proper motions of
the stars.  We  then took a weighted mean, and  found that this result
depended almost  completely on the $\sim$15 galaxies  whose errors are
smallest.
%
\begin{table}
\caption{Data set used in this work.}
\centering \label{obs}
\begin{tabular}{cccc}
\hline\hline  Epoch(date) & EXPTIME & FILT & GO\\
\hline
I (17July2003)&2$\times$1142s$+$4$\times$1185s&F814W&9815 \\
II (13July2005)&2$\times$1200s$+$4$\times$1264s&F814W&10471 \\
\hline
\end{tabular}
\end{table}
%

\begin{figure}[ht!]
       \centering
       \includegraphics[width=7.0cm]{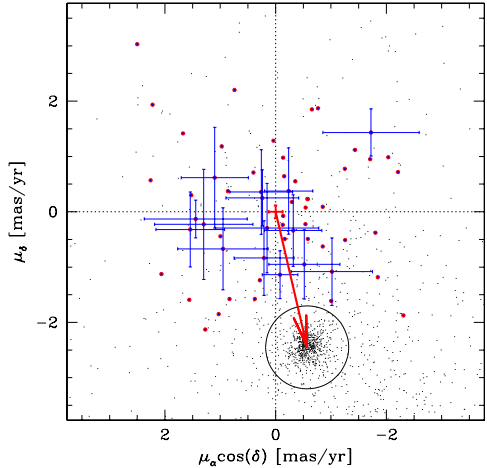}
          \caption{ Vector point diagram of the proper motions, in
          equatorial coordinates.  Filled circles are the reference
          galaxies, and smaller dots are the stars.}
          \label{vector}
       \end{figure}

Figure~\ref{vector}  shows the  vector point  diagram for  the objects
measured.   Dots show  the proper  motions  of the  stars, and  filled
circles those of the galaxies.  To avoid confusion, we show error bars
for only  the best 15  galaxies, each of  which has an error  $<$1 mas
yr$^{-1}$ in  each coordinate.  The  galaxies that have  larger errors
agree with  the mean but contribute  almost no weight  to it.  Cluster
members  form a  tight clump  below the  middle  of Fig.~\ref{vector},
within the  0.75-mas yr$^{-1}$ circle  that we have adopted  to define
cluster membership.   (This radius was chosen as  a compromise between
losing cluster  members and including  field stars that have  a motion
close to that of the cluster stars.)
The separation between field stars  and cluster stars is well defined.
The zero point of the figure was placed at the weighted mean motion of
the galaxies, and is marked with  dotted lines and with the error bars
of  the zero  point.  This  zero-point determination  is  the dominant
source of uncertainty in the absolute motion of NGC 6791.

%
%

\begin{figure}[ht!]
       \centering
       \includegraphics[width=7.0cm]{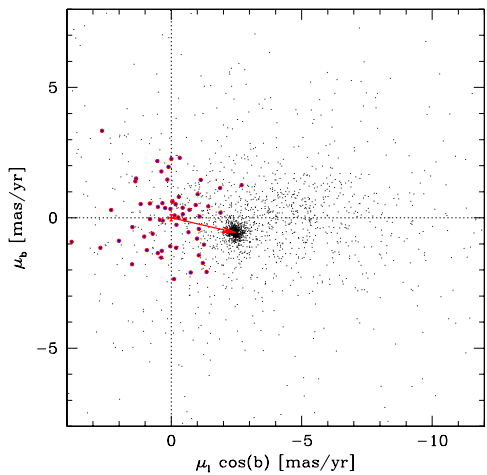}
          \caption{  Vector point  diagram   of  the  proper  motions,  in
          Galactic coordinates.  }
          \label{lb}
       \end{figure}

With this zero  point we find for NGC 6791  an absolute proper motion,
in the J2000.0 system, of
$$(\mu_\alpha \cos\delta,\mu_\delta)=(-0.57,-2.45)\pm(0.13,0.12)~{\rm mas\
yr^{-1}}.$$
These  values represent  a considerable  change from  the $(\mu_\alpha
\cos\delta,\mu_\delta$) $=$ ($-$0.10,  $-$0.70) $\pm$ (0.90, 0.80) mas
yr$^{-1}$  used by Carraro  et al.\  (2006).  In  Galactic coordinates
(Fig.\ \ref{lb}) our motions correspond to
$(\mu_\ell \cos b,\mu_b) = (-2.45,-0.56)$ $\pm$ $(0.13,0.13) ~ {\rm mas\
  yr^{-1}}.$
At the compressed scale of  Fig.~\ref{lb} many more field stars can be
seen.  The distribution of  their proper motions is clearly elongated;
this  is the  consequence  of streaming  effects  due to  differential
rotation, for stars at different distances along the line of sight.
%
%

\section{Radial velocity}
%
For the third  component of the motion of NGC 6791  (along the line of
sight), we used the results  in two recently published papers.  In the
first of these Gratton et al.\ (2006) obtained high-resolution spectra
of four  red-clump stars;  combining their measured  radial velocities
gives a mean value of $V_{\rm rad}$ $=$ $-$47.2 $\pm$ 1.5 km s$^{-1}$.
In  the second paper,  Carraro et  al.\ (2006)  derived a  mean radial
velocity  $V_{\rm rad}$  $=$ $-$47.1  $\pm$ 0.8  km s$^{-1}$  from the
spectra  of 15  probable cluster  members.  We  adopted as  the radial
velocity of NGC 6791 the weighted mean of the two measurements,\\
$ V_{\rm rad} = -47.1 \pm 0.7 ~ {\rm km ~ s^{-1}}. $
\begin{table}[!ht]
\caption{Input conditions for orbit calculation. Distances are in kpc,
and velocities in km s$^{-1}$. }
\centering
\label{orbIN}
\begin{tabular}{cccccc}
\hline\hline
d   & U         & V           & W  ($=$ Z)  & $\Pi$     & $\Theta$    \\
\hline
3.6 & $34\pm 4$ & $-51\pm 2 $ & $-11\pm 2$ & $40\pm 4$ & $168\pm 3 $ \\
4.0 & $38\pm 4$ & $-52\pm 2 $ & $-12\pm 2$ & $43\pm 4$ & $167\pm 3 $ \\
4.4 & $42\pm 5$ & $-54\pm 2 $ & $-13\pm 2$ & $47\pm 4$ & $165\pm 3 $ \\
\hline
\end{tabular}
\end{table}
\begin{table}
\caption{Orbit parameters, for the three different distances.
Units:\
$d${[kpc]},
$L_{\rm z}${[kpc km s$^{-1}$]},
$E_{\rm tot}${[${\rm 10 \times km^2 s^{-2}}$]},
$P ${[Myr]},
$R_{\rm a}${[kpc]},
$R_{\rm p}${[kpc]},
$z_{\rm max}${[kpc]},
$e${[pure number]}.
}
\centering
\label{orbOUT}
\begin{tabular}{cccccccc}
\hline\hline
     $d$ & $L_{\rm z}$  & $E_{\rm tot}$& $P$
                      & $R_{\rm a}$& $R_{\rm p}$  &$z_{\rm max}$&  $e$   \\
\hline
3.6 &  1101        & $-$11085      & 120 &  9.50 & 3.32 & 0.86  & 0.48 \\
4.0 &  1060        & $-$11174      & 130 &  9.83 & 3.09 & 0.98  & 0.52 \\
4.4 &   994        & $-$11232      & 141 & 10.12 & 2.81 & 1.00  & 0.56 \\
\hline                                   
\end{tabular}
\end{table}

%
\section{Calculation of the orbit}
%

Our  absolute  proper motion  for  NGC  6791,  along with  the  radial
velocity and  an assumed distance  of 4 kpc  $\pm$ 10\% (King  et al.\
2005), allows us to derive its three velocity components and calculate
its Galactic orbit.  The orbit  should allow us to study the dynamical
history  of the  cluster, and  to assess  the possible  impact  of the
motion on its internal dynamics, its mass function, and its origin.

The integration of an orbit requires adopting a model of the potential
of the Milky  Way.  We chose that of Allen  \& Santillan (1991), which
assumes a Galactocentric distance and rotation velocity for the Sun of
$R_0 = 8.5$ kpc and $\Theta_0=220$ km s$^{-1}$, and takes densities in
bulge, disk,  and halo  components whose combined  gravitational force
fits  a rotation curve  that agrees  with observation.   Besides being
time-independent, their potential is axisymmetric, fully analytic, and
mathematically very  simple.  It has  already been used to  derive the
Galactic orbits of  open clusters (Carraro \& Chiosi  1994, Carraro et
al.\ 2006) and disk and halo globular clusters (Odenkirchen \& Brosche
1992,     Milone    et     al.\    2006).      The     potential    is
time-independent---clearly   a    crude   approximation,   because   a
significant variation  of the Galactic potential is  expected over the
lifetime of  this cluster.  Nevertheless, it is  reasonable to believe
that the real Galactic potential has  not changed much in the last few
Gyr, so that  the parameters that we derive  for the present-day orbit
of  NGC 6791,  such as  the apo-  and perigalactic  distances,  can be
considered to be reasonable estimates.

\begin{figure}[ht!]
\begin{center}
\includegraphics[width=8cm]{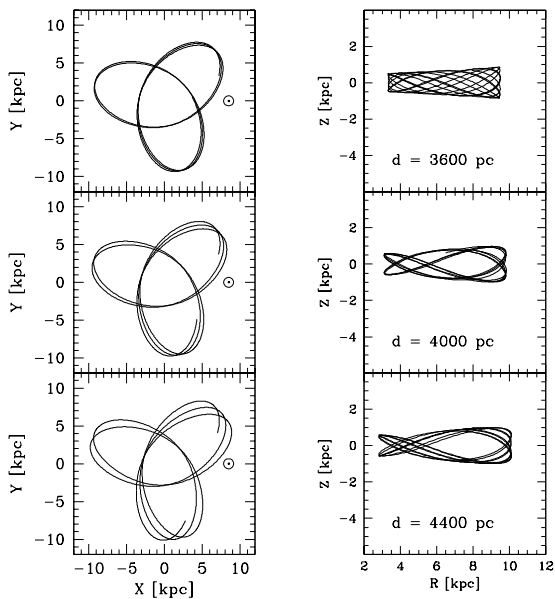}
\protect\caption[]{
Orbits  calculated  back  in  time  for 1  Gyr,  for  assumed  cluster
distances of 3.6, 4.0, and 4.4 kpc.
}
\label{orbs}
\end{center}
\end{figure}

The  initial  conditions are  given  in  Table  \ref{orbIN} for  three
different  heliocentric  distances.   The  integration  routine  is  a
modified second-order Bulirsch-Stoer  integrator (Press et al.\ 1992).
The orbits  were integrated back in time  for 1 Gyr, and  are shown in
Fig.\  \ref{orbs} both in  the $xy$  plane (left  panels), and  in the
meridional plane (right panels).   The orbital parameters are given in
Table  \ref{orbOUT},   where  successive  columns   give  the  assumed
heliocentric distance  of the cluster ($d$), the  $z$-component of the
angular momentum ($L_{\rm z}$),  the total energy ($E_{\rm tot}$), the
orbital period  ($P$), the apo- ($R_{\rm a}$)  and pericenter ($R_{\rm
p}$) of the  orbit, the maximum vertical distance  the cluster reaches
($z_{\rm  max}$)  and  the  eccentricity ($e$),  defined  as  $(R_{\rm
a}-R_{\rm p})/(R_{\rm a}+R_{\rm p})$.

Our  newly derived  orbital parameters  are based  on a  more accurate
space velocity, and are thus  an improvement over previous orbits.  In
particular, this  is the  first time that  proper motions  referred to
extragalactic  objects have  been used  in studying  the orbit  of NGC
6791.

%

The $\pm$10\% range in distance does not change the shape of the orbit
significantly,  nor do  the  orbital parameters  change greatly.   The
orbit is of {\it boxy} type.  The eccentricity is significantly higher
than is typical for an  old open cluster.  As the assumed heliocentric
distance increases  (from top  to bottom panel)  the cluster  tends to
have  a longer  period, to  reach greater  heights above  the Galactic
plane, and to show a larger epicyclic amplitude, dipping closer to the
Galactic Center.  The cluster never moves very far from the Sun toward
the anticenter, while  in the other direction it  reaches rather small
Galactocentric  distances.   Over one  radial  period  it crosses  the
Galactic plane three times.  (This  is quite clear in the bottom panel
on the  right of Fig.\ \ref{orbs};  for other assumed  distances it is
less obvious, but equally true.)

Note that  the new  tangential motion derived  in this paper  leads to
some sizable  differences from  previous results.  The  most important
change  is in  the  apogalactic  distance $R_{\rm  a}$,  which in  the
present  paper  is  strikingly  smaller than  previous  values.   This
weakens significantly  the likelihood  of an extragalactic  origin for
NGC 6791, as proposed by Carraro et al.\ (2006).

In its life in relatively dense regions of the Milky Way, NGC 6791 has
had a  difficult time dynamically.   In each orbital period  of $\sim$
130 Myr it  has endured a rapid Galactocentric passage at  $R \sim $ 3
kpc, a disk crossing at $R \sim  $ 9 kpc, and two more rapid crossings
through the denser part  of the disk at $R \sim $  5 kpc---all four of
them producing  tidal shocks.   The survival of  the cluster  till the
present era is probably due only to its high density and large mass.
The   mass  has   been  decreasing   with  time,   however.   Internal
equipartition keeps  the lower-mass stars preferentially  in the outer
parts,  and  the  tidal   buffeting  has  detached  much  of  low-mass
population, leading  to the flat mass  function noted by  King et al.\
(2005).

Another application  of knowledge of the  orbit of NGC 6791  is to the
question of the origin of  a super-metal-rich cluster that is $\sim$ 8
kpc from  the Galactic center.   Grenon (1999) has suggested  that the
stars in the solar  neighborhood that have comparably high metallicity
could have originated in the Galactic bulge and then been perturbed by
the central  bar into orbits that  bring them out to  here.  Our orbit
suggests that a similar dynamical history might apply to NGC 6791.

%
\section{Summary and Conclusions}
%

By identifying and measuring galaxies with point-like centers, we have
been able to measure an absolute proper motion for NGC 6791.  With the
known radial velocity and the  distance, we have computed the Galactic
orbit of  the cluster.  Uncertainties in  the orbit are  due mainly to
the   inaccuracy  of  the   distance,  but   partly  to   the  unknown
gravitational influence  that the central  bar of the Galaxy  may have
had on the orbit of the  cluster.  It is quite plausible that the high
metallicity  of NGC 6791  is associated  with an  origin in  the inner
region of the Galaxy.

\begin{acknowledgements}
G.\ P.\ acknowledges support by  MIUR under the program PRIN2003.  I.\
R.\  K.\  and J.\  A.\  were supported  by  STScI  grants GO-9815  and
GO-10471.
\end{acknowledgements}

%
%

\end{document}